\documentclass[prb,twocolumn,showpacs,preprintnumbers]{revtex4}
\usepackage{graphicx}
\usepackage{dcolumn}

\begin{document}
{\bf Comment on "Giant Nernst Effect due to Fluctuating Cooper
Pairs in Superconductors"} In a recent Letter, Serbyn et al. [1]
microscopically and phenomenologicaly studied thermomagnetic
effects above the superconducting transition and generalized
results of [2,3] for arbitrary magnetic fields. In our opinion,
the results of [1] disagree with basic physical principles.

(i) In the Gaussian model, using the Kubo method the authors of
[1] calculated the bulk heat current, ${\bf j}^h = T \beta ({\bf
E}\times {\bf H})/H$, and found that the coefficient $\beta$
diverges at $T \rightarrow 0$. To get rid of the contradiction
with the third law of thermodynamics, they amend the heat current
by the "{\it circular magnetization heat current}, ${\bf
j}^Q_M=c({\bf M}\times{\bf E})$," where ${\bf M}$ is the
magnetization. Below we will show that dissipationless
magnetization currents do not transfer the heat and, therefore,
the Kubo method provides an exact expression for the
thermomagnetic coefficient $\beta$.

It is known that in a finite sample, besides the bulk currents
given by the Kubo formulas, charge and energy are also transferred
by surface magnetization currents [4]. Circular electric and
energy magnetization currents,
\begin{eqnarray}
{\bf j}^e_M =
c\nabla \times {\bf M}, \ \ \ {\bf j}^\epsilon_M = \nabla \times
(c \phi {\bf M}) = \phi {\bf j}^e_M + c{\bf M}\times {\bf E}
\end{eqnarray}
($\phi$ is the electric potential), are divergence-free and
corresponding net magnetization currents are always zero (see Eqs.
3, 4, 7 and 39 in [4]). Therefore, instead of adding the surface
magnetization current, one can subtract its bulk counter-flux [4].
Let us consider the surface and bulk magnetization energy currents
in the direction of ${\bf M}\times{\bf E}$ (Fig.1). According to
Eqs. 1, the surface electric magnetization current ${\bf j}^s = c
{\bf M}\times {\bf n}$ (${\bf n}$ is the unit vector normal to the
surface) leads to the surface energy current
\begin{eqnarray}
{\bf J}^\epsilon_s = \phi_A{\bf j}^s_A+\phi_B{\bf j}^s_B,
\end{eqnarray}
 which may be presented as
${J}^\epsilon_s=j^s(\phi_B-\phi_A)=-cMEw$, where $w$ is the width
of the sample. Certainly, the surface energy current ${\bf
J}^\epsilon_s$ and its bulk counter-flux,
\begin{eqnarray}
{\bf J}^\epsilon_b=c({\bf M}\times{\bf E})w,
\end{eqnarray}
are equal and have opposite directions, ${\bf J}^\epsilon_s= -{\bf
J}^\epsilon_b$, as it is shown in Fig. 1.

Note, that Eqs. 2 and 3 do not contain any transport
characteristics. Therefore, they can be derived directly from the
Maxwell equations. To do it, we remind that without magnetization
currents the transformation of electromagnetic energy into the
heat is described by equation
\begin{eqnarray}
- {\rm div} {c {\bf E}\times{\bf H} \over 4\pi} = {\bf
j}^e_{tr}\cdot {\bf E},
\end{eqnarray}
where ${\bf j}_{tr} = c \nabla \times {\bf H} /4 \pi$ is the
transport electric current. Integrating Eq. 4 over the sample
volume, $V$, we get the well-known result: the power dissipated in
the sample, ${\bf j}^e_{tr}\cdot {\bf E} V$, is equal to the
electromagnetic power given by the Pointing vector integrated over
the sample surface.

Now let us add to the above consideration dissipationless
magnetization currents. Taking into account that ${\bf j}^e_M =
c\nabla \times {\bf M}$ and $\nabla \times {\bf E}=0$, we have
\begin{eqnarray}
- {\rm div} ( c {\bf E}\times{\bf M} ) = {\bf j}^e_{M}\cdot {\bf
E}.
\end{eqnarray}
While formally Eq. 5 looks analogous to Eq. 4, it has completely
different physical sense due to the dissipationless nature of the
magnetization currents. To see it, let us consider a sample, where
the magnetization ${\bf M}$ changes in the direction of ${\bf
M}\times{\bf E}$ as it is shown in Fig. 2. We will analyze the
energy balance in a small volume, which is formed by two close
cross-sections, (A,B) and (A`,B`), shifted by $\Delta {\bf R}$ in
the direction ${\bf E}\times{\bf M}$. Integrating the l.h. side of
Eq. 5 over the volume between (A,B) and (A`,B`), we get the net
power carried to the volume by the bulk energy currents, which are
expressed in terms the magnetization-related part of the Pointing
vector,
\begin{eqnarray}
&& - c w \{ [{\bf E}\times ({\bf M}+ \Delta {\bf M})] \cdot {\bf
n}_{A`B`} + [{\bf E}\times {\bf M}] \cdot {\bf n}_{AB}\} \nonumber
\\&&  = \ c w E \Delta M = {\bf J}^\epsilon_b (AB)-{\bf
J}^\epsilon_b (A`B`) = \Delta {\bf J}^\epsilon_b.
\end{eqnarray}
To show that this power is removed by the surface energy currents,
we should present the r.h. side of Eq. 5 in terms of the surface
currents. First, let us note that the change of the magnetization
$\Delta {\bf M}$ between (A,B) and (A`,B`) is created by the
magnetization current $\Delta J^s =c \Delta M$, which flows from
the one side of the sample (A,A`) to another side (B,B') between
the two cross-sections (A,B) and (A`,B`) as it is shown in Fig. 2.
Because of the charge conservation, the current $\Delta J^s$
across the sample is exactly equal to the changes in the surface
magnetization currents between A and A`, and B and B': $\Delta
{J}^s = {j}^s_{B`} - {j}^s_{B} = -({j}^s_{A`} -{j}^s_{A})$.
Finally, integrating the the r.h. side of Eq. 5 over the volume,
we get
\begin{eqnarray}
E \Delta J^s w = (\phi_A - \phi_B) \Delta J^s =\Delta
J^\epsilon_s.
\end{eqnarray}
Eqs. 6 and 7 is nothing more than the integral representation of
l.h. and r.h. sides of Eq. 5. Compare Eq. 6 and 7, we see that in
any volume the bulk energy magnetization currents are compensated
by the surface energy magnetization currents. Obviously, Eqs. 6
and 7 are the finite-difference form of Eqs. 3 and 2
correspondingly.

Thus, in the magnetic field the important part of the energy is
transferred by the surface magnetization currents. To get the net
energy current through the sample, the electromagnetic flux ${\bf
J}^\epsilon_s=-{\bf J}^\epsilon_b=c({\bf E}\times{\bf M})w$ should
be added to the Kubo's energy current [4]. But, as we will see,
for the heat current no such corrections to the Kubo method are
required.

The thermal energy is counted from the electro-chemical potential
$\mu +e\phi$ and the heat current is defined as [5] ${\bf
j}^h={\bf j}^\epsilon - \phi{\bf j}^e - \mu{\bf j}^e/e $. The
surface energy current ${\bf J}^\epsilon_s = \phi_A{\bf
j}^s_A+\phi_B{\bf j}^s_B$ does not have a heat component, because
every $\phi {\bf j}^s$ term in the energy current is canceled by
$\phi {\bf j}^s$ in the definition of ${\bf j}^h$. Naturally, its
bulk counter-flux ${\bf J}^\epsilon_b$ also transfers only
electromagnetic energy ($c{\bf M}\times{\bf E}$ is a magnetization
part of the Poynting vector), which represents reversible work and
can be entirely used. Thus, magnetization heat currents are
absent. This statement also follows from the fact that circular
temperature gradients do not exist (see discussion of Fig. 4 in
[6]) and {\it circular heat currents} requires permanent energy
supply. Finally, contrary to [1] the Kubo method gives an exact
expression for the thermomagnetic heat current [6].

(ii) While for noninteracting electrons, thermomagnetic effects
are proportional to the square of the particle-hole asymmetry
(PHA) and very small, according to [1-3] the fluctuation
thermomagnetic effects do not require PHA and, therefore, huge.
The Gaussian model is fully applicable to ordinary
superconductors, for which the works [1-3] predict the fluctuation
correction to $\beta$ to be at least $\epsilon_F / T \sim 10^5$
times bigger than $\beta$ in the normal state. Certainly, such
huge effects are not known for ordinary superconductors [7]. Also,
the calculations of [1] for superconductors with the negative
interaction constant in the Cooper channel being generalized for
nonsuperconducting metals with a positive constant would also lead
to giant thermomagnetic effects even in ordinary metals.

(iii) The authors of [1] also proposed the phenomenological
theory, where $\nabla T$ was introduced via $\nabla \mu (T(r))$.
The authors claim that in this way they derived a general
Einstein-type relation: $ \nu_N \equiv \beta/(\sigma H)= (\sigma/
ne^2c )(\partial \mu/\partial T)$, where $ \sigma$ is the
electrical conductivity and $n$ is the electron concentration.
However, according to textbooks [5], $\nabla \mu$ should always be
included in the effective electric field and such relation does
not exist. Even with the relation above, to get the giant effect
from the Cooper pairs, the authors of [1] introduce the
thermodynamic chemical potential of pairs in the form:
$\mu_{c.p.}(T)=T_c-T$ [1]. It is known that $\mu_{c.p.}$ is always
zero, because a number of pairs is not conserved.

We note in conclusion, that technically speaking the authors of
[1] calculated $\beta$ for the Aslamazov - Larkin diagram.
Previous works predicted huge thermal and thermoelectric effects
originating from this diagram have been found to be wrong [3].
"The reason is that this diagram corresponds to the contribution
of the superfluid flow to the current. Since the superfluid
carriers no entropy, it does not contribute to the thermal
current" [8].

A. Sergeev, M.Yu. Reizer, and V. Mitin

University at Buffalo, Buffalo, NY 14260.

Fig. 1. Bulk and surface energy currents.

\end{document}